%
%
\input psfig.tex
\magnification=1200
\hoffset=-.25in
\voffset=0.2in
%

\vsize=8.0in
\hsize=6in
\tolerance 10000

\baselineskip 12pt plus 1pt minus 1pt
\centerline{\bf Analytical Estimate of the
Critical Velocity for}
\centerline{\bf Vortex Pair Creation in Trapped Bose Condensates}
\noindent
\vskip .5in
\centerline{\it by}
\medskip
\centerline{\rm M. Crescimanno, C. G. Koay, and R. Peterson}
\smallskip
\centerline{\it Physics Department}
\centerline{\it Berea College}
\centerline{\it Berea, Ky. ~40404}
\medskip
\centerline{\it and}
\medskip
\centerline{\rm R. Walsworth}
\smallskip
\centerline{\it Harvard-Smithsonian Center for Astrophysics}
\centerline{\it 60 Garden Street}
\centerline{\it Cambridge, Ma. ~02138}
\centerline{\it }
\vskip .7in
\centerline{\it December 1999}
\vskip 1.2in
{\bf ABSTRACT :}   We use a modified Thomas-Fermi approximation
to estimate analytically the critical velocity for the formation of
vortices in harmonically trapped
BEC.  We compare this analytical estimate to numerical calculations and
to recent experiments on trapped alkali condensates.

\vfill
\eject

\bigskip
\noindent {\bf I. Introduction :}
The experimental realization of Bose-Einstein condensation (BEC) in cold
trapped alkali atoms has stimulated great interest in weakly interacting
inhomogeneous quantum gases.  The alkali condensates are excellent systems
for the study of quantum fluids, because (i) they are dilute, and hence
admit an accurate description by mean field theory; (ii) system parameters
such as density, temperature, and trapping potential are under fine
experimental control; and (iii) sensitive diagnostics (mostly optical) have
been developed to probe condensate behavior.  In particular, alkali BEC may
serve as a powerful laboratory to study superfluidity -- both
long-established questions about its breakdown, and newer questions about
the effects of finite size and spatial inhomogeneity.

It is well-known theoretically that collective excitations in a weakly
interacting BEC determine a superfluid critical velocity, below which the
condensate flow relative to a perturbing object or potential is without
drag.  This critical velocity is given by the Landau criterion
[1]:
$$ v_c = {\rm min}\biggl({{E(I)}\over{I}}\biggr)
\eqno(1.1)$$
where $E(I)$ is the energy of a condensate collective excitation of linear
momentum (or impulse) $I$.  For liquid $^4$He,
the breakdown of superfluidity is observed
experimentally to occur at a critical velocity that is much smaller than
that due to the excitation of phonons or rotons [2].  Beginning with
Feynman more than 40 years ago [3], it has long been thought that the
critical velocity for superfluid $^4$He is set by the creation of complex
vortex structures (e.g., pairs, rings, and loops) depending on system geometry,
temperature, etc.  However, it has not been possible to fully verify this
basic hypothesis because of the strong interactions in a liquid, which
complicate quantitative comparison of superfluid $^4$He experiments [4] with
theory [5], as well as the limited ability to vary
the liquid density.  Alkali gas BEC may provide an experimental
system to test quantitatively the link between the superfluid critical
velocity and the creation of vortices.

In a landmark recent experiment [6], Ketterle and co-workers observed a
critical velocity for the excitation of an alkali BEC when a perturbing
potential (a blue-detuned laser beam) was moved through the trapped quantum
fluid.  These measurements are in qualitative agreement with both the
well-known analytical calculation for a homogeneous weakly-interacting BEC
in a channel of finite diameter [2,3,7] (providing a $v_c$ that is a factor of
$\sim$
2 lower than experiment), and recent numerical calculations [8-11]
based on the
Gross-Pitaevskii equation [12] (providing a $v_c$ that is a factor of
$\sim$ 2
higher than experiment).

Detailed understanding of the breakdown of superfluidity in Bose
condensates will likely require further experimentation (e.g., to vary the
trapped alkali BEC geometry and density, and to create greater spatial
symmetry in the BEC perturbation), as well as additional theoretical work --
both numerical and analytical.  To this end, we report in this paper an
analytical calculation of the critical velocity for vortex pair production
in a harmonically-trapped dilute Bose condensate.  We employ a
modification of the usual Thomas-Fermi (TF) approximation to the Bogoliubov
mean-field
theory [13], treating the BEC like a
fluid with an exotic equation of state.  We include leading kinetic energy
terms caused by the two counter-rotating vortices, and use these kinetic
energy terms as an effective potential for the BEC.  We then compute,
approximately, the energy $E(I)$ and impulse $I$
of the vortex
pair, and determine the critical velocity from the Landau criterion.

As a step toward checking this modified TF limit for
BEC vortices, we also compute the Bogoliubov many-body
``wavefunction'' ($\psi$) for a
condensate with a single central vortex, and then
calculate $E$, $I$, and $v_c$
for this case.  Since our calculations are based on an
effective hydrodynamic model of BEC,
the vortices that we consider do not, strictly
speaking, have quantized angular momentum in units of $\hbar$.  However, we
find that the modified TF prediction agrees quite well with a
vortex solution computed
numerically.

We believe the advantages of the analytical method presented here are (i)
its simplicity; and (ii) that it works well for the limit of large
inter-particle interaction, which is the limit largely being explored by
current experiments.  In addition, the modified
Thomas-Fermi analysis suggests the manner in which ${v_c}$, as a fraction
of the speed of
sound (${c_B}$),
decreases as the inter-particle interaction increases.

In section II below we introduce both the model used in our calculation,
and the modified
Thomas-Fermi approximation.  In section III we apply this model and
approximation to the creation of vortex pairs near the center of a
harmonically-trapped BEC, and then compare the analytically estimated
critical velocity
with the recent measurements by the MIT group [6].  In section IV we show
that a modified TF analytical calculation of ${v_c}$ for a
single central vortex in a trapped BEC agrees well with numerical
simulation.  Finally, in section V we summarize the implications of this
approach.

\bigskip
\noindent{\bf II. Model of Trapped BEC and the Modified Thomas-Fermi Limit :}
We restrict our discussion to the dynamics of a harmonically-trapped
single-component Bose condensate at $T = 0$, which is well
described by the Gross-Pitaevskii (GP) equation [12]:
$$ -{{\hbar^2}\over{2M}} \bigtriangledown^2 \psi'
+ {{KR^2}\over{2}} \psi' + {U}|\psi'|^2 \psi'
= \mu'\psi'
\eqno(2.1)$$
where the field $\psi'$, the diagonal part of the
multiparticle wave function, serves as the condensate order parameter, and is
referred to as ``the wavefunction of the condensate''; $M$ is the atomic mass;
$\mu'$ is the BEC chemical potential; and $U|\psi'|^2$ is an effective
nonlinear potential arising
from the atomic interactions (due to $s$-wave binary atomic
collisions).  Here
${U} = {{4 \pi \hbar^2 a_s}\over{M}}$, where $a_s$ is the
atomic $s$-wave
scattering length.
For simplicity, we assume the trap has
cylindrical symmetry and take $R^2 = x^2 + y^2 = |{\vec r}~|^2$,
with force constant
$K$ and no trapping potential
along the symmetry (${\hat z}$) axis.
Note that we solve the
above GP equation subject to the constraint that the
total number of atoms
$N$ is fixed,
$$ N = \int~{\rm d}^2r {\rm d}z ~ |\psi'|^2 \qquad .
\eqno(2.2)$$
For ease of calculation we use scaled harmonic oscillator units (h.o.u.) in
which the units of length, time, and energy are
$\sqrt{{{\hbar}\over{2M\omega}}}$
, ${{1}\over{\omega}}$, and
$\hbar\omega$, respectively, where $\omega = \sqrt{ {{K}\over{M}}}$
is the trap angular frequency.
Also, we reduce the GP equation to two dimensions by assuming the
wavefunction is separable: $\psi' = \psi({\vec r}~)\Phi(z)$.
Thus, adopting the notation
of Jackson et al. [10], we find
$$ -\bigtriangledown^2 \psi
+ {{R^2}\over{4}} \psi + C |\psi|^2 \psi
= \mu \psi
\eqno(2.3)$$
where we have scaled $\psi$ so that
$$ 1 = \int~ {\rm d}^2r ~ |\psi|^2 \qquad .
\eqno(2.4)$$
Solving Eq. (2.3) subject to Eq. (2.4)
fixes $\mu$, the dimensionless chemical potential.
The
dimensionless interaction parameter is
$$ C = {{2M {U} N}\over{\hbar^2 a_t \zeta}}
\eqno(2.5)$$
where
$$a_t = \sqrt{ {{\hbar}\over{2M\omega}}}
\eqno(2.6)$$
is the classical turning-point width of the condensate in the harmonic
trap in the $C\rightarrow 0$ limit (equal to the unit of length in h.o.u.),
and
$\zeta$ is the trap aspect ratio given approximately by
$\zeta={{\int {\rm d}z ~z~  \Phi^*\Phi}\over{a_t~|\Phi(0)|^2}}$.

For large $C$ -- which corresponds
to most current experiments -- the solution to the GP equation is well
approximated by the Thomas-Fermi (TF) limit [13].
In this limit one neglects
the derivative term in Eq. (2.3), yielding
$$ \psi(R) = \sqrt{ {{\mu-R^2/4}\over{C}}} \qquad and \qquad
\mu=\sqrt{ {{C}\over{2\pi}}}\qquad ,
\eqno(2.7)$$
where the second relation comes from the normalization
condition, Eq. (2.4). (See also Eq. (3.4) and discussion below.)
Eqs. (2.7) will be used extensively in this paper as the
no-vortex solution that we compare to the vortex solutions. For reference,
note that in the TF limit (i.e., the long wavelength limit),
the local speed of (Bogoliubov) sound
in the condensate is given in h.o.u. by $c_B = \sqrt{2 C |\psi|^2}$
(see for example Ref. [10]).

The many-body wavefunction of a vortex in the condensate will have a
spatially-dependent
phase [see Eq. (3.1) below], and a
spatially-dependent amplitude which we refer to as the 'envelope function'.
Let  $\phi({\vec r}~)$ and $A({\vec r}~)$ be the phase and envelope functions
(both real) parameterizing
this wavefunction via $\psi = A e^{i\phi}$. Using this parameterization
 in Eq. (2.3), and
equating real and imaginary parts, we find:
$$  -\bigtriangledown^2 A + A ({\vec \bigtriangledown} \phi)^2
+ {{R^2}\over{4}} A + C A^3
= \mu A
\eqno(2.8)$$
$$ -\bigtriangledown^2 \phi - 2{\vec \bigtriangledown} A \cdot
{\vec \bigtriangledown} \phi = 0 \qquad .
\eqno(2.9)$$

One may think of Eq. (2.8) as an equation of hydrostatic equilibrium and
Eq. (2.9) as a continuity equation.
{\it What we call the modified TF approximation consists of first
solving the continuity equation
[Eq. (2.9)] for
$\phi({\vec r})$, and then using that
solution in Eq. (2.8) to determine $A({\vec r})$, neglecting
the $\bigtriangledown^2 A$ term.}
Thus, in
this approximation the envelope function $A({\vec r}~)$ is the
solution of a simple algebraic equation.

Once we have normalized the vortex wavefunction via Eq. (2.4), we compute
the energy and
integrated impulse.
The energy functional is
$$ E = \int {\rm d}^2r ~~ \bigl[
{\vec \bigtriangledown} \psi^*\cdot
{\vec \bigtriangledown} \psi
+ {{R^2}\over{4}}\psi^*\psi+{{C}\over{2}}
(\psi^*\psi)^2\bigr]\qquad .
\eqno(2.10)$$
Using $\psi=A e^{i\phi}$ with $A$ and
$\phi$ real, one finds,
$$ E = \int {\rm d}^2r ~~\bigl[({\vec \bigtriangledown}A)^2
+(A{\vec \bigtriangledown}\phi)^2+ {{R^2A^2}\over{4}}
+{{C}\over{2}}A^4\bigr] \qquad .
\eqno(2.11)$$
The local momentum density is ${\vec P} = {{i\hbar}\over{2}}
\bigl(({\vec \bigtriangledown}\psi^*) ~\psi - \psi^*
{\vec \bigtriangledown \psi}\bigr)$, which is used to
compute the total impulse:
$$ I= \int {\rm d}^2r ~|{\vec P}|
$$
$$ \qquad \qquad \qquad = \hbar \int{\rm d}^2r ~ A^2
|{\vec \bigtriangledown}\phi|\qquad .
\eqno(2.12)$$
We can then estimate the critical velocity for vortex pair creation by
applying the Landau criterion [Eq. (1.1)],
$$ v_c = {{E_{vortex}-E_{no~ vortex}}\over{I_{vortex}-I_{no~ vortex}}}
\qquad .
\eqno(2.13)$$
In the next section we describe the details of an analytical computation of
this ratio for a vortex pair near the center of the trap.

\bigskip
\noindent{\bf III. Analytical Calculation of the Vortex Pair Critical
Velocity :}
We consider a dilute BEC containing a pair of counter-rotating vortices
with opposite charge or vorticity
$\pm n$. The vortices are assumed to be located symmetrically about the
trap center, with the cores a
distance
$d$ apart.   We apply
the modified Thomas-Fermi (TF)
method to the hydrostatic equilibrium expression
[Eq. (2.8)] by assuming that $\bigtriangledown^2 A$ is negligible.
In addition, we adopt
the ansatz that the two vortices are far enough from each other, but near
enough to the trap center, that the total phase advance about the vortex pair
is the simple sum of the phase advances of the two vortices viewed
individually.  This ansatz is akin to the
analytical approximation developed by Feynman [3] for homogeneous BEC,
and also used by Fetter in a similar context [14].  Practically, the
ansatz is equivalent to requiring ${{1}\over{\mu}} < d^2 < \mu$
in dimensionless harmonic
oscillator units (h.o.u.).  Operationally it means that the wavefunction's
phase $\phi({\vec r})$ satisfies $\bigtriangledown^2\phi = 0$
everywhere outside the vortex cores,
which implies from the continuity expression [Eq. (2.9)] that
${\vec \bigtriangledown A}\cdot
{{\vec \bigtriangledown} \phi}$ must be zero in
this region.  Our solution for $A({\vec r})$ in this case
satisfies the continuity equation near the vortex cores, and also far away
from the vortices, since radial gradients of the phase vanish as
${{1}\over{R^2}}$.  In
addition, the spatial integral of the continuity equation vanishes
identically, and deviations due to the
approximations described above are
asymptotically bounded.

We have found that neglect of the $\bigtriangledown^2 A$
term in Eq. (2.8) leads to the
main systematic errors in the analytical calculation described here.
However, the deviations are small for parameters typical of current
experiments.

With the approach outlined above, the phase of the vortex pair wavefunction is
$$ \phi({\vec r}) = n~ {\rm atan}
\biggl[{{\sin\theta - {{d}\over{2R}}}\over{\cos\theta}}
\biggr] - n~ {\rm atan}
\biggl[{{\sin\theta + {{d}\over{2R}}}\over{\cos\theta}}\biggr]
\eqno(3.1)$$
where $R$ and $\theta$ are the usual polar coordinates as
measured from the trap
center.  Employing the modified TF method, we find the condensate
wavefunction envelope to be
$$ A^2({\vec r}) = {{1}\over{C}} \biggl(\mu-{{R^2}\over{4}}
-{{d^2n^2}\over{R^2d^2\cos^2\theta + (R^2-d^2/4)^2}} \biggr)
\eqno(3.2)$$
where the last term is the leading kinetic energy contribution, due to the
$|{\vec \bigtriangledown} \phi|^2$ term in Eq. (2.8).
Note that this kinetic energy contribution
is simply ${{1}\over{(r_1r_2)^2}}$,
where $r_1$ and $r_2$ are the respective distances
measured from the vortex cores to the point $(R,\theta)$.

The above calculation scheme for $A({\vec r})$ breaks down very close to
the vortex cores (small $r_1$ or
$r_2$) and also at large $R$, since the right hand side of
Eq. (2.2) becomes negative. Considering these regions to be excluded
complicates the analytic evaluation
of energy and impulse  [Eqs. (2.11) and (2.12)] precisely where the TF
approximation fails.
To circumvent this calculational difficulty
we adopt a regulated expression for $A({\vec r})$:
$$ A^2({\vec r}) = {{1}\over{C}}
\biggl(\mu-{{R^2}\over{4}}
-{{d^2n^2}\over{R^2d^2\cos^2\theta
+ (R^2-d^2/4)^2 + \epsilon}} \biggr)
\eqno(3.3)$$
where $\epsilon = {{d^2n^2}\over{\mu}}$.
This regulated expression is justified by the
observation that for vortex pairs not too far from the trap center
($d^2 < \mu$), the contribution to
$A({\vec r})$ from the kinetic energy term $|{\vec \bigtriangledown} \phi|^2$
is never larger than $\mu$.
The regulation markedly affects the wavefunction
envelope near the vortex cores, but has a
small effect on the
calculated energy, impulse, and critical velocities.  (This is shown for a
trap-centered, single vortex in section IV below.)  In sum, the use of
the regulated expression for $A({\vec r})$ is an important
practical step in our
analytical calculation, because it allows us to perform the spatial
integrals for $E_{vortex}$ and $I_{vortex}$ over the entire plane,
rather than excluding regions near the vortex cores.

We begin the integrations by normalizing the condensate wavefunction for
the cases of no vortices and a single vortex pair. Referring to Eq. (2.7), the
normalization for the no vortex
case yields
$$ \mu \equiv \mu_0 = \sqrt{{{C}\over{2\pi}}}\qquad ,
\eqno(3.4)$$
where we have performed the integration in Eq. (2.4) out to a maximum radius,
$R_{TF} = 2\sqrt{\mu}$, i.e., the Thomas-Fermi condensate edge
(the radius at which the
condensate chemical potential is dominated by the trap potential in the
Thomas-Fermi limit).  Similarly, inserting Eq. (3.3) in Eq. (2.4), and
noting that $|\psi|^2 = A^2$, we determine the normalization
condition for the vortex pair wavefunction:
$$ {{C}\over{2\pi}} = \mu^2-n^2 {\rm ln}
\biggl({{1+{{\mu d^2}\over{n^2}}}\over{1+{{d^2}\over{8\mu}}}}\biggr)
\eqno(3.5)$$
where we have again performed the spatial integration out to the
Thomas-Fermi condensate edge, which for large $C$ is
$R_{TF} = 2\sqrt{\mu} - {{n^2d^2}\over{16\mu^{5/2}}}+\ldots$.
Note that
Eq. (3.5) indicates that $\mu$ increases with the vortex charge, $n$, as
expected.  (See Ref. [15] for details of the calculation of these
normalization conditions.)

Next, we compute the condensate energy.  For the case of no vortex, we insert
Eq. (2.7) in Eq. (2.10), and find
$$ E_0 = {{4\pi\mu_0^3}\over{3C}} \qquad .
\eqno(3.6)$$
For a vortex pair of charge ${\pm n}$, we insert in Eq. (2.11) both the
vortex pair's
phase given in Eq. (3.1) and the regulated expression for $A({\vec r})$
given in
Eq. (3.3).  Performing the spatial integration again out to
$R_{TF}$ (see [15] for details),
we find the energy for a vortex pair to be:
$$ E_n = {{\pi}\over{2C}}\bigl[\mu^2R_{TF}^2-{{R_{TF}^6}\over{48}}
\bigr] -{{d^2n^2\pi}\over{8C}}\biggl[ {{8\mu^2}\over{1+d^2\mu/n^2}}
+{\rm
ln}\biggl({{R_{TF}^4\mu^2(1+d^2/2R_{TF}^2)}\over{n^4(1+d^2\mu/n^2)}}\biggr)
\biggr]\qquad .
\eqno(3.7)$$
Using Eqs. (3.6) and (3.7), we compute the energy difference between a trapped
condensate with a symmetric pair of  unit-charged, counter-rotating ($n=1$)
vortices, and the no
vortex ground state ($n=0$).
To leading order in the limit of large $C$,
we find
$$ E_{1} - E_{0} =
{{2\pi\mu_0}\over{C}}{\rm ln}\bigl({{d^2\mu_0+1}}\bigr)
+\ldots
\eqno(3.8)$$
Note that the energy difference vanishes in the $d \rightarrow 0$ limit,
as expected.
Also, the energy difference varies as the natural logarithm of the distance
between the vortices (at large separation), which resembles results from
earlier studies of vortices in homogeneous BEC [2].

Next, we calculate the impulse $I_n$ for the condensate with a symmetric
pair of
counter-rotating
vortices with charge $\pm n$.  (With no vortices, the trapped condensate
impulse is zero.)  Using Eq. (3.1) we find
$$ |{\vec \bigtriangledown} \phi| =
{{nd}\over{[R^2d^2\cos^2\phi + (R^2-d^2/4)^2]^{{1}\over{2}}}}\qquad .
\eqno(3.9)$$
We insert this formula and the regulated expression for
$A^2({\vec r})$ [Eq. (3.3)]
into Eq. (2.12) to determine the vortex pair impulse.  We find the largest
contribution to the spatial impulse integral comes at large $R$, and is thus
dependent on the condensate size (set roughly by the Thomas-Fermi
condensate edge, $R_{TF}$) as well as the vortex pair charge magnitude ($n$)
and separation
($d$).  Evaluating the integral out to
$R_{TF}$, and assuming large $C$, we find
the vortex pair impulse to be
$$ I_{n} = {{nd\mu_0\pi}\over{C}}{\rm ln}
\bigl( {{64\mu_0}\over{ed^2}}
\bigr)+\ldots
\eqno(3.10)$$
to leading order (see [15] for details). Note that like
the energy, the impulse for vortex pair creation vanishes in the small $d$
limit, and is proportional to the vortex charge, as expected.  Also, since
we are using a macroscopic hydrodynamic model in our calculations, the
quantization of angular momentum of the BEC wavefunction does not lead to a
simple quantization condition on the integrated impulse.

Assembling the results for the energy and impulse leads directly to an
estimate of the critical velocity at which symmetrically-placed,
oppositely-charged vortex pairs, at relative separation $d$, can form near
the center of a Bose condensate trapped harmonically in two dimensions.  In
the large $C$ limit by using Eq. (2.13) we find
$$v_c \simeq {{2}\over{d}}{{\ln(d^2\mu_0)}\over{\ln({{64\mu_0}\over{ed^2}})}}
\eqno(3.11)$$
where again, $\mu_0 = \sqrt{{C}\over{2\pi}}$. These expressions indicate a
weak dependence (via $\mu_0$
and $C$ within the logarithms) of $v_c$ on the BEC non-linear self-energy.
Recall that the
derivation of Eq. (3.11) assumes that
$1/\mu < d^2 <
\mu$ in dimensionless harmonic oscillator units (h.o.u.).  The first
inequality is the requirement that
the vortices be at least several healing lengths away from each other, and
the second
inequality is the requirement that they are not too far from the trap
center.

These conditions seem to be largely met in the recent MIT critical velocity
experiment using
a trapped sodium BEC [6], although this experiment and our model have
different spatial symmetries
(i.e., geometries). Onset of energy dissipation in the MIT condensate was
observed at a
critical velocity when moving a repulsive barrier (a blue-detuned laser
beam) through the middle of the
cold atom cloud. The laser beam was directed along the smaller, radial
direction of the cigar-shaped
condensate, which had Thomas-Fermi diameters of 45 $\mu$m and 150 $\mu$m in
the radial and axial
directions, respectively.  The laser barrier was moved back-and-forth along
the condensate's axial
direction (perpendicular to the axis of the laser beam) at a constant
speed, and energy dissipation was
determined from measured changes in the condensate fraction. Thus the
experiment was explicitly a
three-dimensional system with anisotropic trapping in the plane
perpendicular to the laser beam axis.

Ignoring the different geometry of our two-dimensional,
isotropically-trapped BEC model, we use the
analytical calculation outlined above to estimate the critical velocity for
vortex pair creation in
the MIT condensate.  We assume the vortex cores are parallel to the axis of
the blue-detuned laser beam,
and are separated by a distance $d$ equal to the laser beam diameter (13
$\mu$m). In the central region
of the MIT condensate, the number of atoms per unit length parallel to the
laser beam axis ($n_L$) was
about 3 x 10$^{9}$ cm$^{-1}$. Using 2.9 nm for the sodium $s$-wave
scattering length ($a_s$), we find
from Eq. (2.5) that $C \simeq  8\pi a_s n_L \simeq  25,000$ (in h.o.u.) for
the MIT BEC experiment.

Using these values in  Eq. (3.11), we estimate that the onset of
vortex pair formation, and hence
energy dissipation,
occurs at a critical velocity $v_c
\simeq $ 0.4 mm/s, a factor
of about four below the measured value [6].  (Note: this estimated $v_c$
varies by $\sim 30\%$ over
the range of trap frequencies present in the MIT experiment -- from 18 Hz
in the axial direction to 65
Hz in the radial direction.) We defer further interpretation of this
estimate of $v_c$ to the
conclusion.

\vskip .2in
\noindent{\bf IV. A Test of the Modified Thomas-Fermi Limit :}
As a test of the analytical calculation of vortex critical velocity using
the modified Thomas-Fermi (TF) approximation, we consider a simple, symmetric
example: a single vortex of charge $n$ at the center of an isotropic BEC
that is harmonically-trapped
in two dimensions.  In this case, the cylindrical symmetry of the system
requires the phase of the
condensate to advance proportional to the polar angle $\theta$, and hence
Eq. (2.9) limits
the envelope function $A({\vec r})$ to be a
function of $R =|{\vec r}|$ alone.
Also, the single-valuedness of the condensate
wavefunction constrains the phase to be $\phi = n\theta$, where $n$
is an integer
(the vortex charge).  Therefore, Eq. (2.8) reduces to a second order
ordinary differential equation for $A(R)$ that can be integrated
numerically, {\it without making the modified
Thomas-Fermi approximation}, which can then be used to test the modified TF
analytical calculation.  We
employed an adaptive mesh relaxation algorithm to perform the numerical
integration
of Eq. (2.8) for
both $n = 0$ and $n = 1$, using the value $C = 200,000$
for the coefficient of the non-linear term throughout.
The computed $A(R)$ for $n=0$ is shown in
Figure 1 and compared to the corresponding $A(R)$
calculated
analytically using the traditional TF approximation. 
\medskip
\centerline{\psfig {figure=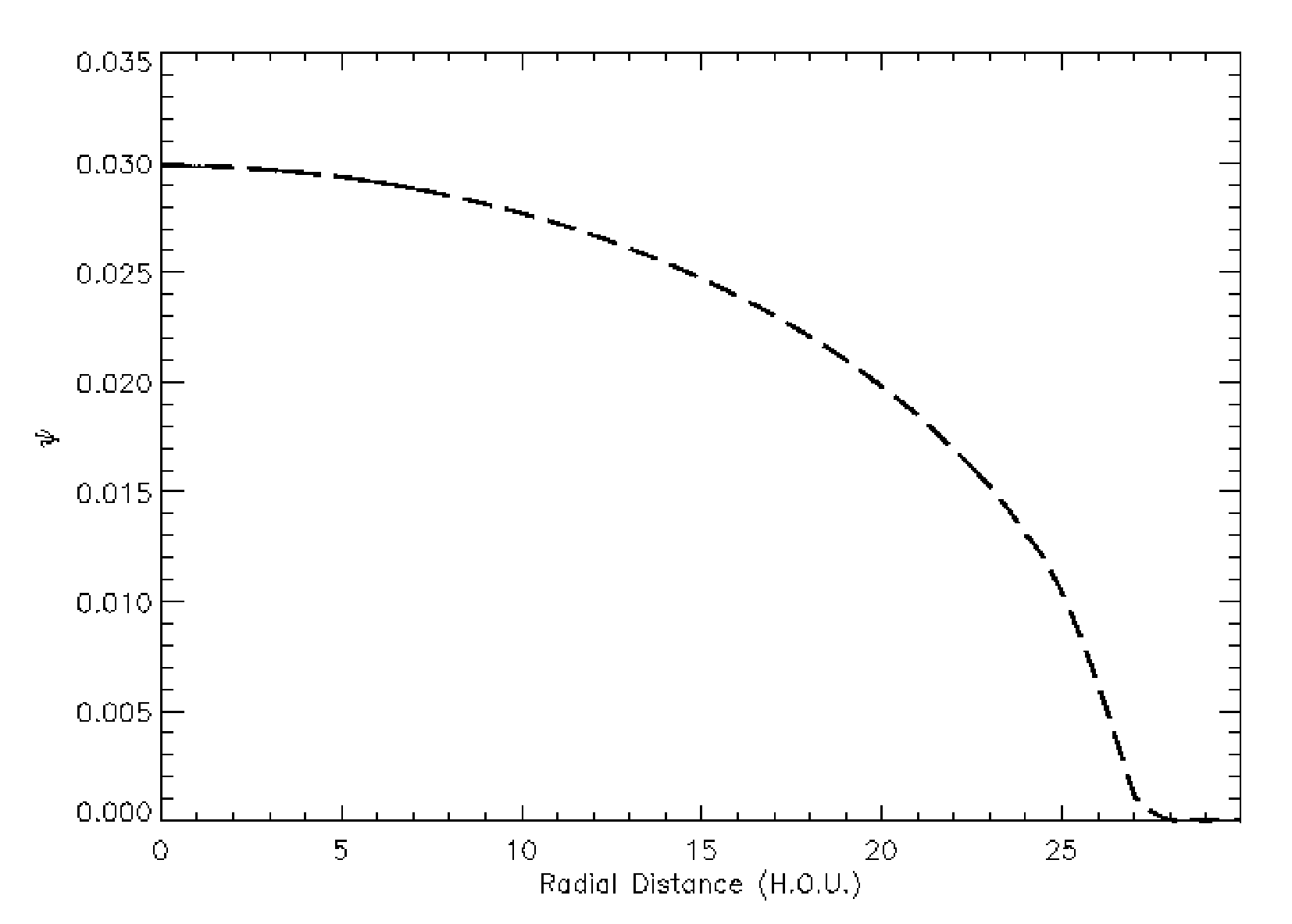,height=3.0in,angle=270}}
\medskip
\centerline{\it Fig. 1:  Calculated envelope function $A(R)$ for a condensate}
\centerline{\it with no vortex ($n = 0$).  Numerical and TF analytical}
\centerline{\it calculations give identical results on the scale of this
graph.}
\bigskip
Figure 2
provides a comparison of the numerical and modified TF analytical solutions
for $A(R)$
with a single central vortex of charge $n = 1$.
\medskip
\centerline{\psfig {figure=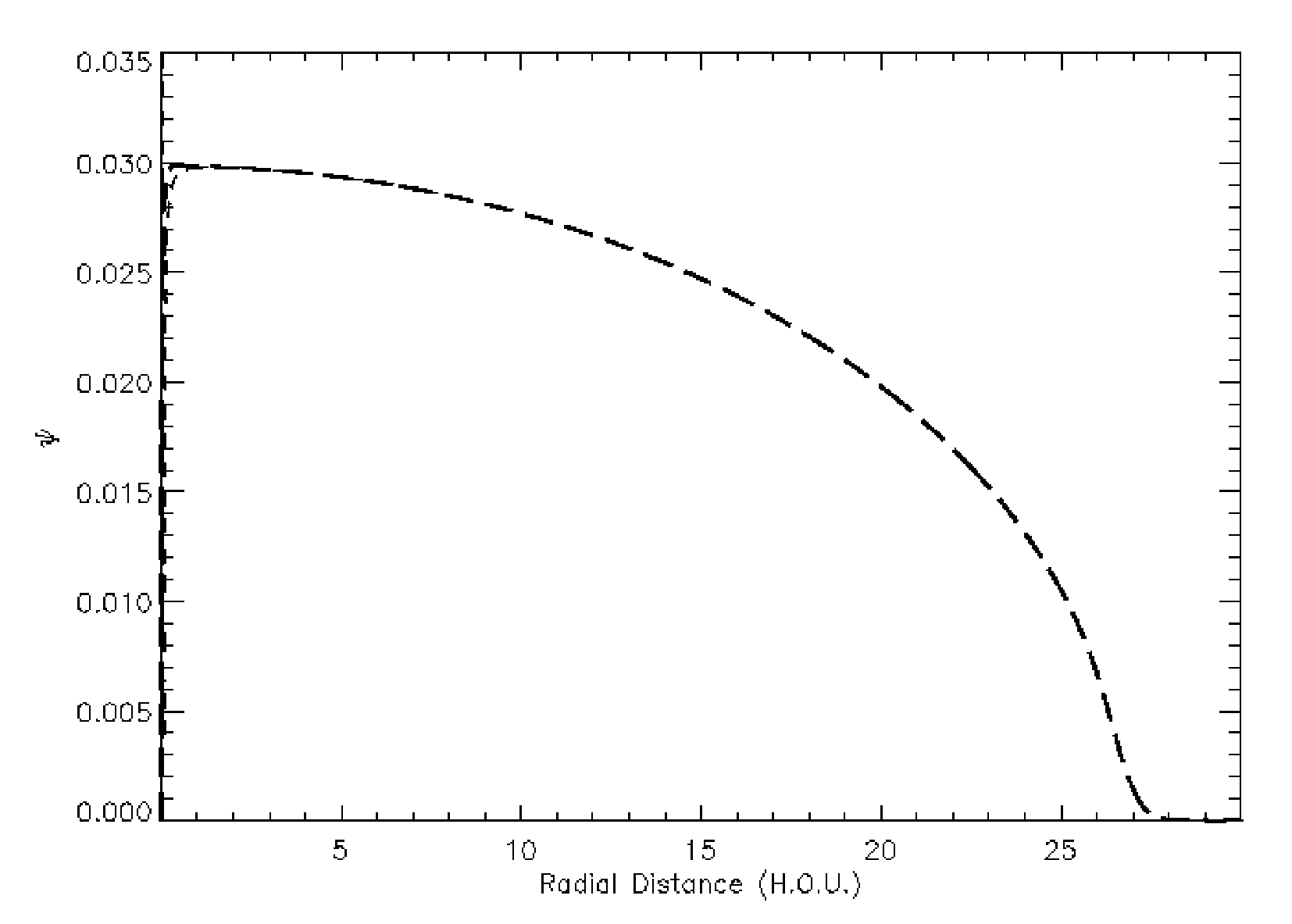,height=2.0in,angle=270}
\psfig {figure=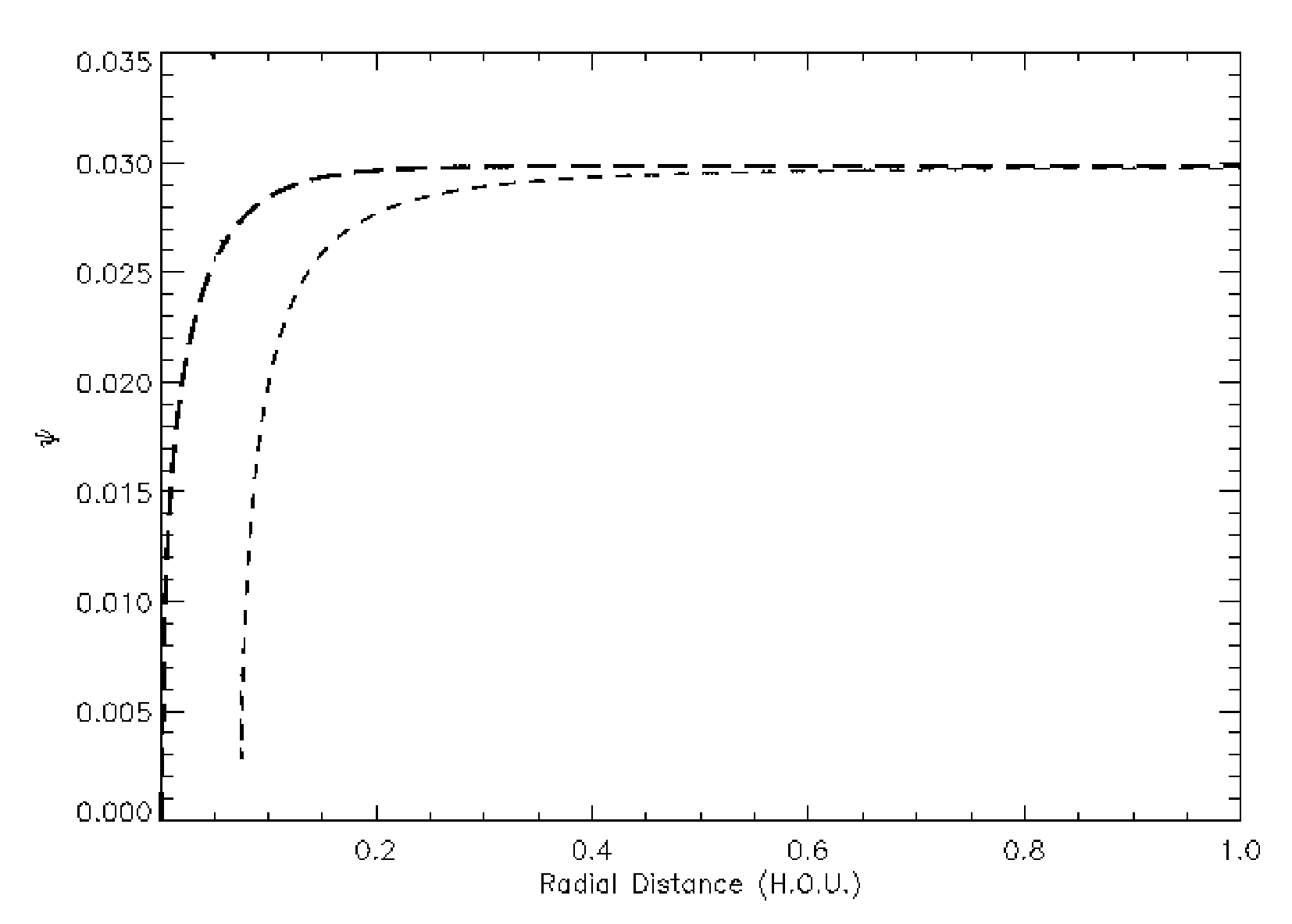,height=2.0in,angle=270}}
\medskip
\centerline{\it Fig. 2:  Comparison of numerical and modified
TF envelope functions $A(R)$}
\centerline{\it for a condensate with a single central vortex of unit charge
$(n = 1)$.}
\centerline{\it (a) Full radial extent of $A(R)$: differences not
perceptible on this scale.}
\centerline{\it (b) $A(R)$ near the trap center.  Dashed line is modified
TF calculation.}
\medskip
In the modified TF limit, the analytical envelope function for a
single central vortex of charge $n$ is
$$ A(R) = \sqrt{ { {\mu-{{R^2}\over{4}}-{{n^2}\over{R^2}}}\over{C}}}
\qquad , \qquad \phi(\theta) = n\theta \qquad .
\eqno(4.1)$$
This expression for $A(R)$ shows that including the
leading kinetic energy term due to the vortex in the analytical calculation
is equivalent to an
effective potential barrier near the
vortex core caused by the angular momentum of the condensate.
In this approximation the condensate extends over the annulus
$R_0< R < R_{max}$, outside of which
$A$ vanishes. From Eq. (4.1) we see that
$$ R_0^2 = 2\mu-2\sqrt{\mu^2-n^2} \qquad and \qquad
R_{max}^2 = 2\mu+2\sqrt{\mu^2-n^2} \qquad .
\eqno(4.2)$$
Hence in the modified TF limit, the vortex core radius (in h.o.u.) is
effectively $n/\sqrt{\mu}$ at large $\mu$.
On general grounds, we expect the radius
of the vortex core to be of the order of the condensate healing length,
$\xi = {{1}\over{8\pi a_s n_0}}$,
where $a_s$ is the $s$-wave scattering length and $n_0$ is the local density.
Thus the modified Thomas-Fermi limit reproduces the expected scaling of
vortex core size, since $\xi \sim {{1}\over{\sqrt{\mu}}}$
in dimensionless units.  Note also
that at large $R$ (ignoring the trap potential) the effect of the vortex on
the condensate envelope
falls off as $1/R^2$ in the modified TF approximation, which is identical
to  the asymptotic behavior
of the trial vortex wavefunction in homogeneous BEC used by Fetter
[14].

The modified TF analytical calculation of the
energy of a single
central vortex of charge $n$ is
found by using Eq. (4.1) in Eq. (2.11) (ignoring the
${\vec \bigtriangledown} A$ term) and integrating
over the annulus given in Eq. (4.2):
$$ E_n = {{4\pi (\mu^2-n^2)^{{3}\over{2}}}\over{3C}} \qquad .
\eqno(4.3) $$
[Compare with Eq. (3.6).] Although not obvious from this equation, $E_n$
increases as a
function of $n$ due to the dependence of $\mu$ on $n$
through the normalization condition $\int {\rm d}^2r ~|\psi|^2 = 1$.
In particular, we find this normalization requires
$$ {{C}\over{2\pi}} = \mu \sqrt{\mu^2-n^2}
-{{n^2}\over{2}} {\rm ln} \biggl( {{\mu + \sqrt{\mu^2-n^2}}
\over{\mu-\sqrt{\mu^2-n^2}}}\biggr) \qquad ,
\eqno(4.4)$$
hence causing $R_{max}$ (Eq. (4.2)) to increase with $n$.
Next, using Eq. (4.1)
in the analytical calculation of the impulse
[Eq. (2.12)]
of a single
central vortex, we find
$$ I_n = {{8\pi n}\over{3 C}} (\mu-n)^{{3}\over{2}} \qquad .
\eqno(4.5)$$
For the no-vortex case ($n = 0$) and $C = 200,000$, numerical
integration of Eq. (2.3) yields the total condensate energy: $E_0 = 118.965$.
[All numbers here are in harmonic oscillator units (h.o.u.).]  With the
analytical method we find $E_0 = {{4\pi\mu_0^3}\over{3C}} = 118.94$,
where $\mu_0 = \sqrt{{C}\over{2\pi}} = 178.4$.
For a single central vortex with $n = 1$ and $C = 200,000$, the numerical
solutions give
$E_1 = 118.991$ and $I_1 = 0.0996$, whereas the modified TF
analytical calculation yields $E_1 = 118.97$ and $I_1 = 0.0989$. Using
these values, we
find $v_{c} = {{\Delta E}\over{\Delta I}} = 0.26$ (numerical)
and 0.30 (analytical). [Recall that the impulse integral is identically zero
for the no vortex solution.] This
reasonable agreement suggests that the modified Thomas-Fermi approximation
in the large $C$ limit should enable analytical calculations of 10-20\%
accuracy for vortex critical velocities in inhomogeneous BEC, provided the
calculational model and
experiment have the same spatial symmetries.

\medskip
\noindent{\bf IV: Conclusion :}
In this paper we report an analytical calculation of the critical velocity
for creation of a pair of counter-rotating vortices in
harmonically-trapped,
dilute
Bose condensates.  Excitation of
vortices may set the lowest limit for a superfluid critical velocity in
BEC.  For the analytical calculation presented here, we
developed a modified Thomas-Fermi approximation to the standard mean-field
theory [13].  This approximation is
appropriate to the limit of large interparticle interactions, and thus
is relevant to most current experiments.  The approximation consists of two
steps, described above in Section II: (i) solve a continuity equation for
the phase of the condensate wavefunction; and then (ii) insert this
solution for the phase in an equation of hydrostatic equilibrium and solve
for the condensate envelope function, ignoring the second order spatial
derivative of the envelope function.  In essence, we include contributions
to the condensate kinetic energy that come from gradients of the phase (a
signature of vortices), but assume that vortex-induced gradients in the
condensate envelope function are negligible.  Alternatively, one can think
of the modified Thomas-Fermi approximation as equivalent to treating
BEC like a fluid with an exotic equation of state, with the leading vortex
kinetic energy terms acting as
an effective potential. Once the envelope function is determined, we
compute approximately
the energy and integrated linear momentum (or impulse) of the vortex pair,
and determine the
critical velocity from the Landau criterion, $v_c = min(energy/impulse)$.

As described in Section III, we find qualitative agreement between our
analytical calculation and the
BEC excitation critical velocity recently measured by Ketterle and
co-workers at MIT [6].  This result
is not surprising given the obvious difference in symmetry between our
model and the MIT critical velocity experiment: we assume the vortex core
axes to be parallel to the axis of cylindrical symmetry in a
two-dimensional trap confining the condensate; whereas, the experimental
excitation would likely create vortices with core axes perpendicular to the
finite
cylindrical axis in the actual, quasi-two-dimensional trap.
In addition, vortices produced at
velocities substantially below
the observed critical velocity may collectively dissipate energy at a rate
comparable to the decay rate of the unstirred condensate itself, making
it difficult to identify $v_c$ experimentally (i.e., the MIT experiment may
only place an upper limit
on $v_c$).  We expect that
detailed understanding of the role of vortex creation in the breakdown of
superfluidity in BEC will require further experimental development to
enable more symmetric excitation of the condensate, as well as theoretical
treatments (both numerical and analytical) of more realistic systems.
Nevertheless, as shown in Section IV, the modified Thomas-Fermi analytical
method
is in reasonable agreement with numerical calculations
based on the full GP equation
of the critical velocity
{\it for a vortex with high symmetry}: specifically, a
single vortex in the center of a two-dimensional
harmonic trap.

We note two important issues that are not included in the analysis
presented in this paper: (i) the breakdown of the hydrodynamic
approximation at the edges of the trapped condensate; and (ii) the dynamics
of vortex excitation in the presence of nominally smooth trapping and
perturbing potentials.  Further theoretical and experimental work is
necessary on these problems.

We conclude by emphasizing that the modified Thomas-Fermi approximation
presented in this paper is straightforward to implement.  In addition to
enabling an analytical calculation of vortex critical velocity
(for more details see Ref. [15]) this approximation should
have other applications in the
physics of trapped BEC.

We thank E. Timmermans for helpful discussions.

\vfill
\eject
\ \
\vskip .3in
\centerline{\bf References}
\vskip .3in
\item{1.} See, for example, I. M. Khalatnikov, {\it Introduction to the
Theory of Superfluidity}
(Benjamin, New York, 1965).
\item{2.} A nice discussion of the breakdown of superfluidity in liquid
$^4$He, including many important
references, is given in: R.K. Pathria, {\it Statistical Mechanics}
(Pergamon Press, Oxford, 1972).
\item{3.} R.P. Feynman, in {\it Progress in Low Temperature Physics}
(North-Holland, Amsterdam, 1955),
Vol. 1, p. 17.
\item{4.} A. Amar, Y. Sasaki, R.L. Lozes, J.C. Davis, and R.E. Packard,
Phys. Rev. Lett. {\bf 68}, 2624
(1992); O. Avenel and E. Varoquaux, Phys. Rev. Lett. {\bf 55}, 2704 (1985).
\item{5.} P.W. Anderson, Rev. Mod. Phys. {\bf 38}, 298 (1966); E.R.
Huggins, Phys. Rev. A {\bf 1}, 332
(1970); L.J. Campbell, J. Low Temp. Phys. {\bf 8}, 105 (1972).
\item{6.} C. Raman, M. K\"ohl, R. Onofrio, D.S. Durfee, C.E. Kuklewicz, Z.
Hadzibabic, and W.
Ketterle, Phys. Rev. Lett. {\bf 83}, 2502 (1999).
\item{7.} A.L. Fetter, Phys. Rev. Lett. {\bf 10}, 507 (1963); M.P. Kawatra
and R.K. Pathria, Phys. Rev.
{\bf 151}, 132 (1966).
\item{8.} T. Frisch, Y. Pomeau, and S. Rica, Phys. Rev. Lett. {\bf 69},
1644 (1992).
\item{9.} C. Huepe and M.-E. Brachet, C.R. Acad. Sci. Paris {\bf 325}, 195
(1997).
\item{10.} B. Jackson, J.F. McCann, and C. S. Adams, Phys. Rev. Lett. {\bf
80}, 3903 (1998).
\item{11.} T. Winiecki, J.F. McCann, and C. S. Adams, Phys. Rev. Lett. {\bf
82}, 5186 (1999).
\item{12.} V.L. Ginzburg and L.P. Pitaevskii, Sov. Phys. JETP {\bf 7}, 858
(1958); E.P. Gross, J. Math.
Phys. {\bf 4}, 195 (1963).
\item{13.} See discussion in the review article: F. Dalfovo, S. Giorgini,
L.P. Pitaevskii, and S.
Stringari, Rev. Mod. Phys. {\bf 71}, 463 (1999).
\item{14.} A.L. Fetter, Phys. Rev. {\bf 138}, A429 (1965).
\item{15.} M. Crescimanno and R. Walsworth, to appear.

\vfill
\par
\end